\documentclass{PoS}

\usepackage{graphicx}

\providecommand{\href}[2]{#2}

\title{%
\vspace{-7ex}\begin{minipage}[t][1pt][r]{\textwidth}
\begin{flushright}
\rm\small
FR-PHENO-2010-09,
IPPP/10/08,
LAPTH-002/10,
Nikhef/2010-003,
TTK-10-13
\end{flushright}
\end{minipage}\\[6ex]
NLO Cross Sections for the LHC using GOLEM: Status and Prospects
}

\ShortTitle{NLO Cross Sections for the LHC using GOLEM: Status and Prospects}

\author{Thomas Binoth\\
	The University of Edinburgh, School of Physics,
	Edinburgh EH9\,3JZ, UK
      }
\author{Gavin Cullen\\
	The University of Edinburgh, School of Physics,
	Edinburgh EH9\,3JZ, UK\\
	E-mail: \email{g.j.cullen@sms.ed.ac.uk}}
\author{Nicolas Greiner\\
	Institute for Theoretical Physics,
	University of Z\"urich,
	CH-8057 Z\"urich,
	Switzerland\\
	E-mail: \email{greiner@physik.uzh.ch}}
\author{Alberto Guffanti\\
	Physikalisches Institut,
	Albert-Ludwigs-Universit\"at Freiburg,
	79104~Freiburg~i.~Br,
	Germany\\
	E-mail: \email{alberto.guffanti@physik.uni-freiburg.de}}
\author{Jean-Philippe Guillet\\
	LAPTH,
	Annecy-le-Vieux 74951, France\\
	E-mail: \email{guillet@lapp.in2p3.fr}}
\author{Gudrun Heinrich\\
	IPPP,
	Department of Physics,
	University of Durham,
	Durham DH1\,3LE,
	UK\\
	E-mail: \email{gudrun.heinrich@durham.ac.uk}}
\author{Stefan Karg\\
	Institut f\"ur Theoretische Physik~E,
	RWTH Aachen,
	52056~Aachen, Germany\\
	E-mail: \email{karg@physik.rwth-aachen.de}}
\author{Nikolas Kauer\\
	Department of Physics,
	Royal Holloway,
	University of London,
	Egham TW20\,0EX, UK\\
	School of Physics and Astronomy,
	University of Southampton,
	Southampton SO17\,1BJ, UK\\
	E-mail: \email{n.kauer@rhul.ac.uk}}
\author{\speaker{Thomas Reiter}\\
        Nikhef, 1098\,XG Amsterdam, The Netherlands\\
        E-mail: \email{thomasr@nikhef.nl}}
\author{J\"urgen Reuter\\
	Physikalisches Institut,
	Albert-Ludwigs-Universit\"at Freiburg,
	79104~Freiburg~i.~Br,
	Germany\\
	E-mail: \email{juergen.reuter@physik.uni-freiburg.de}}

\abstract{%
In this talk we review the GOLEM approach to one-loop calculations
and present an automated implementation of this technique.
This method is based on Feynman diagrams and an advanced reduction of
one-loop tensor integrals which avoids numerical instabilities.
We have extended our one-loop integral library \texttt{golem95}
with an automated one-loop matrix element generator to compute
the virtual corrections of the process $q\bar{q}\rightarrow b\bar{b}b\bar{b}$.
The implementation of the virtual matrix element has been interfaced
with tree-level Monte Carlo programs to provide the full result
for the above process.}

\FullConference{RADCOR 2009 - 9th International Symposium on Radiative Corrections 
  (Applications of Quantum Field Theory to Phenomenology) \\
  October 25-30 2009\\
  Ascona, Switzerland}

\begin{document}

\section{Overview}

The successful start of the LHC and its first data taking mark the dawn
of a new era in particle physics. Very soon we will be able to explore
the energy scale of electro-weak symmetry breaking which enables us to
either confirm the Standard Model as a low energy theory of particle physics
or to discover new particles guiding us to extensions of the Standard Model.
No matter what the outcome of the experiment will be, a successful interpretation
of the data will require a large number of predictions calculated at least
to Next-To-Leading Order~(NLO) in QCD. Some of these reactions involve up to
four particles in the final state~\cite{:2008uu,Bern:2008ef,Buttar:2006zd,Campbell:2006wx}.

The computation of NLO corrections to processes involving many final state particles 
are cumbersome and time consuming. The time required for a single calculation 
--- without the use of automated programs --- often coincides with the duration of a Ph.D.
Considering the fast experimental progress it is very likely that already
within the next two or three years the physics results of the LHC will
raise the demand for precision calculations within and beyond the Standard
Model for many multileg processes. Only the automatisation of NLO calculations will allow 
us to keep pace with the requirements set by the experiments.

We argue that the automatisation of the computation of NLO cross-sections
will also improve the possibility of comparing results from different
implementations, especially if the tools are made public and common 
conventions are used for input/output and interfacing to external programs are in use. 
A first step in this direction has been made through the Binoth Les Houches Accord~\cite{Binoth:2010xt}. 
This accord exploits the modular structure of NLO calculations and proposes the
reflection of this structure in the implementation of such calculations
as computer programs.

Any QCD cross-section at NLO can be written in the form
\begin{equation}
\sigma_{2\rightarrow N}^{\mathrm{NLO}}=
   \sigma_{2\rightarrow N}^{\mathcal B}
 + \sigma_{2\rightarrow N}^{\mathcal V}
 + \sigma_{2\rightarrow N+1}^{\mathcal R}.
\end{equation}
The first term on the right-hand side describes the Born-level
cross-section calculated from the squared tree-level amplitude
of the $2\rightarrow N$ process. The virtual corrections
$\sigma_{2\rightarrow N}^{\mathcal V}$ stem from the interference term
between tree-level and one-loop diagrams. The third term describes the
real radiation of an extra, unobserved parton at tree-level.

The last two terms lead to singularities which can be regularised
by introducing a non-integer dimension $n=4-2\varepsilon$, yielding
poles in $1/\varepsilon$ which cancel only after both terms have been
added up. For practical applications it is therefore
convenient to introduce subtraction terms, such that
both $\sigma_{2\rightarrow N}^{\mathcal V}$ and
$\sigma_{2\rightarrow N+1}^{\mathcal R}$ are finite and can be
integrated over phase space~independently.

A full implementation of an NLO calculation can therefore be modularised into
a Monte-Carlo integrator for phase space integration,
a tree-level matrix element generator for
	$\sigma_{2\rightarrow N}^{\mathcal B}$ and
	$\sigma_{2\rightarrow N+1}^{\mathcal R}$,
a one-loop matrix element generator for
	$\sigma_{2\rightarrow N}^{\mathcal V}$ and
infrared subtraction terms.
Typically the first two of these components are implemented in the same program.
An overview
of existing techniques and recent contributions to all of the modules can
be found, for example, in~\cite{Bern:2008ef} and~\cite{Binoth:2010xt}.

Methods for computing~$\sigma_{2\rightarrow N}^{\mathcal V}$
can be classified in two groups.
On the one hand there are unitarity based methods determining the coefficients
of scalar one-loop integrals by exploiting analyticity of the
amplitude; on the other hand there are Feynman diagrammatic techniques
starting from tensor integrals, which are then reduced to simpler integrals
which can be evaluated in a numerically stable way. Both techniques have led
to automated implementations in recent years (\cite{Ellis:2007br,%
Ossola:2007ax,Berger:2008sj,Giele:2008bc,Lazopoulos:2008ex,Winter:2009kd} and
\cite{Hahn:1998yk,Kurihara:2002ne,Belanger:2003sd,Bredenstein:2010rs,%
Binoth:2008uq}
respectively).

In Sections~\ref{golem_sec:golem95} and~\ref{golem_sec:golem20} we
present the automated implementation of a method using Feynman diagrams for the 
calculation of~$\sigma_{2\rightarrow N}^{\mathcal V}$, the GOLEM method, based on 
the reduction scheme proposed in \cite{Binoth:2005ff}.
This implementation has been used to compute the NLO virtual corrections
to $q\bar{q}\rightarrow b\bar{b}b\bar{b}$, which is one of the two partonic initial
states contributing to the $pp\to b\bar{b}b\bar{b}$ process.
This process constitutes an important background for Higgs searches
in models beyond the Standard Model where Higgs bosons decay
predominantly into $b$-quarks as discussed in~\cite{Lafaye:2000ec,Krolikowski:2008qa}.

\section{The Generic Integral Form Factor Library \texttt{golem95}}
\label{golem_sec:golem95}

The Feynman diagrammatic approach to computation of one-loop corrections to processes
with N external particles requires the evaluation of tensor integrals of rank $r$ which
 have the general form
\begin{equation}
  I_{\mathsf{N}}^{n;\mu_1\ldots\mu_r}(a_1,\ldots,a_r;S)=
  \int\!\frac{\mathrm{d}^nk}{i\pi^{n/2}}\frac{q_{a_1}^{\mu_1}\cdots%
    q_{a_r}^{\mu_r}}{(q_1^2-m_1^2+i\delta)\cdots
    (q_{\mathsf{N}}^2-m_{\mathsf{N}}^2+i\delta)},
\end{equation}
where $q_i=k+r_i$ and $S$ denotes the matrix $S_{ij}=(r_i-r_j)^2-m_i^2-m_j^2$. 

It is well-known~\cite{Melrose:1965kb} that such a tensor integral can be expressed 
in terms of basis scalar integrals, but the reduction procedure introduces inverse Gram 
determinants ($\det G$) in the coefficients of the expansions which can lead 
to numerical instabilities in certain regions of the phase space.

Therefore, in~\cite{Binoth:1999sp,Binoth:2005ff} we have proposed a
reduction scheme which allows to write any N-point amplitude as a linear
combination of basis integrals ($I_2$, $I_3^n$, $I_3^{n+2}$, $I_4^{n+2}$, 
$I_4^{n+4}$) with and without Feynman parameters in the numerator,
avoiding the introduction of inverse Gram determinants. 

The evaluation of the basis functions can be performed by reducing them further to
scalar integrals using recursion formulae. This further reduction, however, 
introduces Gram determinants in the coefficients which could lead to numerical
instabilities in certain regions of the phase space. 

Potentially dangerous regions are identified by the criterion
$\vert\det G\vert < \Lambda\vert\det S\vert$ for a fixed cut-off~$\Lambda \sim 10^{-5}$.
In this regions the basis integrals are evaluated numerically without applying any
further reduction.

For all integrals without internal masses we have worked out
one-dimensional integral representations
which can be evaluated by numerical integration.

This algorithm has been implemented in form of a Fortran\,90 library,
\texttt{golem95}, for massless internal propagators~($m_i=0$)~\cite{Binoth:2008uq}. 
This version of the code has been made available for download\footnote{%
\href{http://lappweb.in2p3.fr/lapth/GOLEM/golem95.html}{%
\tt http://lappweb.in2p3.fr/lapth/GOLEM/golem95.html}}.

We have recently extended the library \texttt{golem95} to the case
where internal masses are present. 
All infrared divergent integrals have been implemented explicitly. 
For evaluating the finite boxes and triangles the user needs to link the 
LoopTools library~\cite{Hahn:1998yk,Hahn:2000kx,vanOldenborgh:1989wn}. 
This ``massive'' version of the \texttt{golem95} library is currently 
in the testing phase and will be available shortly.

For integrals with internal masses, the option to evaluate the tensor
integrals numerically prior to reduction 
in regions where the Gram determinant tends to zero, is not yet 
supported. 
However, one-dimensional integral representations valid for all 
possible kinematic configurations are under~construction.

\section{Implementation of a One-Loop Matrix Element Generator}
\label{golem_sec:golem20}
Building upon \texttt{golem95} as a loop-integral library, our next step
was the construction of a matrix-element generator at the one-loop level.
The computation is carried out projecting the amplitude onto helicity and colour structures. 
The virtual corrections can therefore be expressed as
\begin{equation}
  \mathrm{d}\sigma_{2\rightarrow N}^{\mathcal V}=
  \frac{1}{n_an_b}\sum_{\{\lambda_i\},j,k}\!\!\!
  {\mathcal{A}_j^{\mathcal B}(p_a^{\lambda_a},p_b^{\lambda_b};
    p_1^{\lambda_1},\ldots,p_N^{\lambda_N})}^\dagger
  \left\langle c_j\vert c_k\right\rangle
  {\mathcal{A}_k^{\mathcal V}}(p_a^{\lambda_a},p_b^{\lambda_b};
  p_1^{\lambda_1},\ldots,p_N^{\lambda_N}) + h.c.
\end{equation}
where $p_i^{\lambda_i}$ denotes the pair of momentum $p_i$ and
helicity label $\lambda_i$ of the $i-$th particle.
The matrix $\left\langle c_j\vert c_k\right\rangle$ consists of
the contractions of all colour basis tensors for a given process
evaluating to rational functions in the number of colours $N_C$
and the normalisation constant $T_R$ of the generators.
The constants $n_a$ and $n_b$ represent the averaging over spin and~colour.

The one-loop amplitude $\sum_k \mathcal{A}_k\vert c_k\rangle$ consists of
a sum of Feynman diagrams, which we generate using QGraf~\cite{Nogueira:1991ex}. 
We use QGraf together with \LaTeX{} and Axodraw~\cite{Vermaseren:1994je} also
for drawing the diagrams; the layout of the diagrams is determined using
the algorithm of~\cite{Ohl:1995kr}.

The expressions of the diagrams are then processed using Form~\cite{Vermaseren:2000nd} and 
the Form library \texttt{spinney} which we have developed for dealing with helicity 
spinors and $n$-dimensional Dirac and Lorentz algebra efficiently. 
Majorana spinors can also be dealt with thanks to the the implementation the flipping rules
for spin lines as described in~\cite{Denner:1992vza}. 

At the moment the GOLEM program can import CompHep~\cite{Boos:2004kh} model files to 
perform Beyond the Standard Model computations. An interface to FeynRules~\cite{Christensen:2008py} 
is under construction.

After the Form program has decomposed the diagram expression into colour structures and the 
tensor integrals are represented in terms of integral form factors as defined in 
\texttt{golem95}~\cite{Binoth:2008uq}, the resulting expressions are optimised and translated into 
Fortran\,90 functions using the code generator \texttt{haggies}~\cite{Reiter:2009ts}. 
At this step the number of multiplications is minimised applying a Horner scheme and
common subexpression elimination.

The generated Fortran\,90 program is linked with \texttt{golem95} for
the numerical evaluation of the tensor integral form factors.
A future version of the program will support the Binoth Les Houches
Accord~\cite{Binoth:2010xt} to facilitate the interfacing to
Monte-Carlo~generators.

\section{NLO Results for $q\bar{q}\rightarrow b\bar{b}b\bar{b}$ \cite{Binoth:2009rv}}
\label{golem_sec:results}

The setup described in the previous section has been used to
compute the virtual corrections of the QCD corrections to
$q\bar{q}\rightarrow b\bar{b}b\bar{b}$ in the limit $m_b=0$ and
$m_t\rightarrow\infty$. We have compared the results with an
independent implementation using FeynArts and FormCalc~\cite{Hahn:1998yk}
to generate and simplify the diagrams, where the tensor integrals 
are algebraically reduced to scalar integrals using the procedure
described in~\cite{Binoth:2005ff}.

In order to compute $q\bar{q}\rightarrow b\bar{b}b\bar{b}$ at NLO accuracy
the Born level cross-section, the real emission contribution and the 
infrared subtraction terms also need to be evaluated.
Since we are only interested in a process with 4 tagged $b$-jets the relevant 
process for the real emission contribution is $q\bar{q}\rightarrow b\bar{b}b\bar{b}g$.

We have used \texttt{MadGraph/MadEvent}~\cite{Maltoni:2002qb,Alwall:2007st} 
and \texttt{MadDipole}~\cite{Frederix:2008hu} to evaluate the tree-like
contributions and to perform the phase space integration.
As an alternative setup, based on an extended version of the 
\texttt{Whizard}~\cite{Kilian:2007gr,Moretti:2001zz} Monte Carlo generator
with an implementation of infrared subtraction terms has been~used to 
obtain an independent cross-check.
For the infrared subtraction we have used Catani-Seymour
dipoles~\cite{Catani:1996vz}
in both implementations including
a phase space slicing parameter following~\cite{Nagy:2005gn}.

To define a $b\bar{b}b\bar{b}$ event, we first we apply a $k_T$ jet
algorithm~\cite{Blazey:2000qt} to decide if the gluon should be
merged with a $b$-quark. If the gluon is merged we use the effective
momentum $\tilde{p}_b=p_b+p_g$ as the momentum of the $b$-quark in the
cuts and in the observables.
Then we apply a $p_T$ cut of $p_T(b_j)>30\,GeV$ and a rapidity cut of
$\vert\eta\vert<2{.}5$ to all
$b$-quarks and a separation cut of $\Delta R>0{.}8$ to all pairs
of~$b$-quarks.

We sum over $q\in\{u,d,c,s\}$ and use the CTEQ6M parton distribution
functions~\cite{Pumplin:2002vw} with two-loop running of $\alpha_s$
for both Leading and Next-to-Leading Order computations. 
The centre of mass energy is set to $\sqrt{s}=14\,GeV$. In our results we use
a fixed factorisation scale of $\mu_F=100\,GeV$; the renormalisation
scale we set to~$\mu_0=\sqrt{\sum_j p_T^2(b_j)}$.

\begin{figure}[hbtp]
\begin{center}
\includegraphics[width=0.8\textwidth]{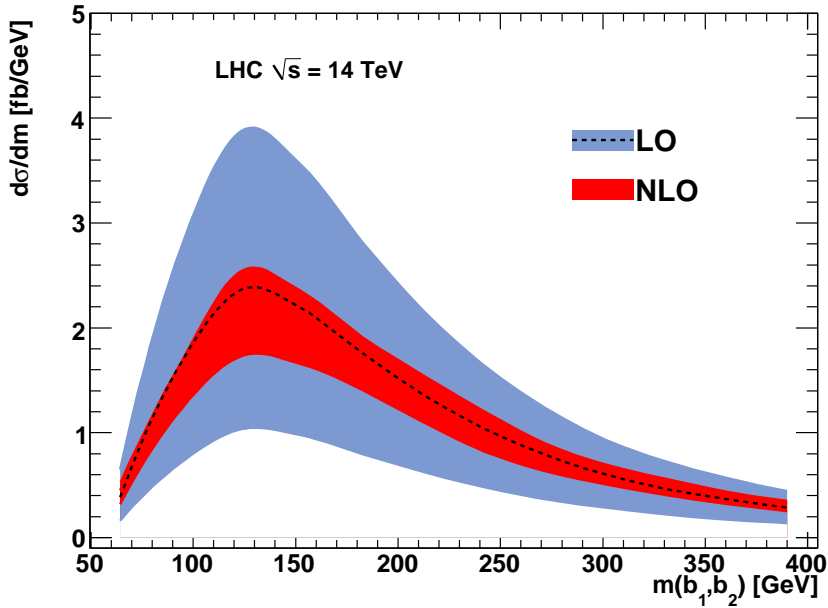}
\end{center}
\caption{Invariant mass ($m_{bb}$) distribution of the two leading
$b$-quarks (see text). The error bands are obtained by varying
the renormalisation scale $\mu_R=x\mu_0$ between $1/4<x<2$, where
$\mu_0=\sqrt{\sum_j p_T^2(b_j)}$. The dashed line shows the
value of the leading order prediction for $x=1/2$.}\label{fig:golem_mbb}
\end{figure}

In Figure~\ref{fig:golem_mbb} we show the invariant mass distribution
of the system of the two $b$-pairs with highest $p_T$. The error bands
have been obtained by varying the renormalisation scale $\mu_R=x\mu_0$
between $1/4<x<2$. The dashed line marks the leading order distribution
for $x=1/2$, which turns out to be very similar to the NLO prediction
for this value. The reduction of the uncertainty band due to scale variations 
clearly shows the importance of the NLO corrections for the precision of
this calculation.

\section{Conclusion}
We have presented results obtained using the GOLEM method and recent
progress in the implementation of an automated one-loop matrix element
generator. A first important step towards this goal was the development
of a one-loop integral library, \texttt{golem95}, which is currently being
extended to the case of massive propagators. As a second step we have
added a completely automated framework which generates efficient
Fortran\,90 code for the numerically stable evaluation of one-loop
matrix elements from a set of Feynman rules. This framework includes
the development of new tools such as an optimising code generator
(\texttt{haggies}), and a Form library (\texttt{spinney}) for the treatment
of helicity spinors and $n$-dimensional Dirac and Lorentz algebra.
Finally, we have presented the complete NLO result for the process
$q\bar{q}\rightarrow b\bar{b}b\bar{b}$, which is a subprocess of
$pp\rightarrow b\bar{b}b\bar{b}$, an important background to
Higgs searches beyond the Standard Model.

In the near future we plan to implement the interface to Monte-Carlo tools
described in~\cite{Binoth:2010xt} and to make all parts of the program
publicly available.

\end{document}